\documentclass[journal, dvipsnames, fleqn]{IEEEtran}

\usepackage{graphicx, xcolor, here}
\usepackage{cite}
\usepackage{amsmath}
\usepackage{amsfonts}
\usepackage{amssymb}
\usepackage[linesnumbered,ruled,vlined]{algorithm2e}
\usepackage{textcomp}
\usepackage{xcolor}

\newcommand{\hi}[1]{\textbf{\color{MidnightBlue} #1}}

\usepackage{tikz}     % Core TikZ package
\usetikzlibrary{arrows,shapes,positioning,calc,backgrounds,decorations.markings,quotes,angles}

\title{A Framework and a \texttt{Python-}Package for Real-Time NMPC parameters Setting}
\author{Mazen Alamir\thanks{The author is with Univ. Grenoble Alpes, CNRS, Grenoble INP, GIPSA-lab, 38000 Grenoble, France. Email: mazen.alamir@grenoble-inp.fr. Website: https://www.mazenalamir.fr.}}
\date{\today}
\begin{document}

\maketitle

\begin{abstract}
This paper presents a framework that enables a systematic and rational choice of NMPC design components such as control updating period, down-sampling period for prediction, control parameterization, prediction horizon's length, the maximum number of iterations as well as penalties on the terminal cost and the soft constraints. The rationale that underlines the design choices is based on real-time implementability, convergence and constraints satisfaction for a given computational device and a specific optimization algorithm. Moreover, a freely available associated \texttt{Python}-based implementation is also described with a fully developed illustrative example implementing a nonlinear MPC controller for a Planar Vertical Take-Off and Landing (PVTOL) aircraft under control saturation and state constraints. 
\end{abstract}

\begin{IEEEkeywords}
Nonlinear Model Predictive, Real-Time implementation, Certification, Nonlinear Systems, Python Package. 
\end{IEEEkeywords}

\section{Introduction}
It is needless to say that Nonlinear Model Predictive Control (NMPC) \cite{Mayne2000, rawlings2017model} is currently the most effective and widely used feedback design methodology in \textit{academic} works that address the control of constrained nonlinear systems. The theoretical foundations of NMPC are now quite established. Moreover free and easy-to-use programming frameworks \cite{Andersson2018} embedding multiple efficient and trustworthy\footnote{As far as sufficient regularity conditions hold.} dedicated solvers \cite{biegler2009large, houska2011acado} for the NMPC-underlying optimization problems are now available. Nevertheless, practitioners still \hi{lack a systematic design procedure for the NMPC components}\footnote{See later for a more precise definition.} that helps achieving the very last step, namely the real-time implementation on a specific computation target device. \\ \ \\
Addressing real-time implementability of NMPC feedback is by now a quite \textit{old} topic. The real-time iteration, first proposed in \cite{alamir2001nonlinear} to address a specific application problem and then generalized by \cite{diehl2002real, diehl2005real, gros2020linear} already enabled a huge extension of the number of successful applications to \textit{fast systems}. While these works suggest performing a single optimization step per sampling period, later works \cite{alamir2009framework, alamir2013monitoring, alamir2016state, bonne2017experimental} suggested, including through experimental implementation, that in general, the appropriate choice of the control updating period in NMPC must ideally be viewed as a state-dependent decision that is tightly linked to many factors such as the dynamics, the implementation hardware, the solver and its associated certification\footnote{This is not the certification of the NMPC closed-loop behavior. It is only the link between the number of iterations of an algorithm and the precision of the corresponding sub-optimal solution. This is only a single item in  the whole NMPC certification as explained in \cite{alamir2015certification,alamir2016state}.} rules \cite{richter2011computational, pu2016complexity}, the level of model discrepancies as well as the prediction horizon's length to cite but a part of the parameters involved \cite{alamir2016state}. Notice however that the above attempts only consider the way the solution of a \textbf{given formulation} is \textit{distributed on-line} over the real lifetime of the process.
\begin{center}
\begin{tikzpicture}
\node[rounded corners, fill=Black!5, inner xsep=2mm, inner ysep=4mm, draw=Blue](O){\color{Blue}
\begin{minipage}{0.42\textwidth}
 The present paper addresses the different problem of \textit{choosing off-line}, for a given target computation device and a given optimization algorithm, the different major ingredients of NMPC formulation so that convergence, constraints satisfaction and real-time implementability are enhanced.
\end{minipage} 
};
\node[rounded corners, fill=white, draw=MidnightBlue] at(O.north){\scriptsize The problem addressed in the paper};
\end{tikzpicture}
\end{center} 
Beside the above rationale determination of the NMPC setting's parameters, the paper briefly describes a freely available\footnote{\texttt{https://github.com/mazenalamir/MPC\_tuner}} \texttt{Python}-based package that implements the proposed solution. \\ \ \\
The paper is organized as follows: The problem under consideration is first clearly described in Section \ref{sec_problem_statement}. The implementation of the optimization algorithm that \textit{tunes} the parameters of the NMPC setting is discussed in Section \ref{sec_theAlgo}. A brief description of the \textit{Python} package that implements the algorithm is proposed in Section \ref{sec_package} and a concrete example showing its use is depicted in Section \ref{sec_example}. Finally the paper ends with a conclusion and a brief discussion regarding possible extensions of the package scope and utilities. 
\section{Problem Statement}\label{sec_problem_statement}
\noindent In order to clearly state the problem under consideration, we need to define the items to be fed as inputs to  the algorithm (Section \ref{sec_inputalgo}) and those NMPC parameters that need to be \textit{tuned} by the algorithm and hence represent its delivered output (Section \ref{sec_outputalgo}). 
\subsection{The input items to the algorithm}\label{sec_inputalgo}
\begin{itemize}
\item[$\checkmark$] A set of \hi{Ordinary Differential Equations} (ODEs) governing the dynamics of the system to be controlled which takes the form $\dot x=f(x,u,p)$ where $x\in \mathbb X\subset \mathbb R^{n_x}$, $u\in \mathbb U\subset \mathbb R^{n_u}$ and $p\in \mathbb P\in \mathbb R^{n_p}$ represent the state, control input and model's parameter vectors respectively. These vectors belong to the compact \textbf{hypercubes} $\mathbb X$, $\mathbb U$ and $\mathbb P$ respectively. \vskip 1mm
\item[$\checkmark$] \hi{A basic sampling period $\tau>0$} which is tailored such that the associated \texttt{Runge-Kutta} (RK) integration scheme yields accurate forward prediction should the expression of $f$ be perfectly known. Since real-time implementability is our main concern, $\tau$ should ideally be the largest sampling period meeting the above requirements. \vskip 1mm
\item[$\checkmark$] \hi{A stage cost and a terminal penalty} functions denoted respectively by $\ell(x,u,p, q)$ and $\Psi(x,p,q)$ in which $q\in \mathbb R^{n_q}$ is a vector of task-related parameters such as set-point values, environment-induced bounds on some components of the state\footnote{which impact the cost function via the exact penalties on the soft constraints violation.}, estimated values of some exogenous parameters impacting the definition of optimality, etc. These \textbf{problem-dependent} maps are used in the definition of the NMPC cost function. \vskip 1mm
\item[$\checkmark$] \hi{A constraint map} of the form $c(x,u,p,q)\le 0\in \mathbb R^{n_c}$ that defines the constraints to be enforced at each instant of the prediction horizon. Notice that all the constraints other than the input saturation-related ones are treated as \textbf{soft constraints}\footnote{This is the only option to yield a robust outliers-proof implementation of constrained NMPC feedback laws.} through appropriate highly weighted exact penalty (to be tuned as shown later on). The constraints satisfaction is enforced through a certification procedure. \vskip 1mm
\item[$\checkmark$] \hi{Some bounds on the search domain} that are detailed later on.
\item[$\checkmark$] \hi{A random sampling function} that can be used to generate representative cloud of initial states $x_0$, model's parameters vector $p$ and context parameter vector $q$.
\end{itemize}
\subsection{The NMPC parameters to be tuned (Algorithm's output)}\label{sec_outputalgo}
To be more specific, the following NMPC ingredients are determined by the design procedure: \vskip 1mm
\begin{itemize}
\item[$\checkmark$] \hi{The control updating period $\tau_u=\kappa \tau$} expressed as an integer multiple $\kappa$ of the basic sampling period $\tau$ invoked above. This explicitly means that the computation rounds that update the control profiles and hence the implemented closed-loop control are separated by $\kappa\tau$ time units. The next figure illustrates this parameter for $\kappa=3$. \vskip 4mm
\begin{center}
\begin{tikzpicture}
\draw[rounded corners, fill=Black!5, draw=white] (0.35,-0.6) rectangle (8.3,1.8);
\def\z{1}
\draw[->, >=latex] (0.5,0) -- node[below, pos=0.98]{\scriptsize \texttt{Time}} (8,0);
\draw[->, >=latex] (0.5,0) -- (0.5,1.4) node[right]{\footnotesize \color{MidnightBlue} MPC open-control profile for $\kappa=3$.};
\foreach \x in {1,...,14}
\filldraw (0.5*\x,0)  circle (1pt);
\draw[thin, MidnightBlue] (0.5,0) -- +(0,\z);
\draw[thin, MidnightBlue] (1,0) -- +(0,0.3*\z);
\draw[thin, MidnightBlue] (2,0) -- +(0,\z);
\draw[thin, MidnightBlue] (3.5,0) -- +(0,\z);
\draw[thin, MidnightBlue] (5,0) -- +(0,\z);
\draw[thin, MidnightBlue] (6.5,0) -- +(0,\z);
%-----
\draw[very thick, MidnightBlue] (0.5,0.8) -- (2,0.8);
\draw[very thick, MidnightBlue] (2,0.4) -- (3.5,0.4);
\draw[very thick, MidnightBlue] (3.5,0.6) -- (5,0.6);
\draw[very thick, MidnightBlue] (5,0.1) -- (6.5,0.1);
%-----
\draw[<->] (2,-0.2) -- node[MidnightBlue, midway, below]{\footnotesize $\tau_u=3\tau$}(3.5,-0.2);
\draw[<->, MidnightBlue] (0.5,0.2\z) -- node[midway, above]{\scriptsize $\tau$}(1,0.2\z);
%----
\end{tikzpicture}
\end{center} 
\item[$\checkmark$] \hi{The prediction's precision parameter $\mu_d\in [0,1]$} that determines the precision of the integration used in the MPC-related computation. More precisely, if no precision drop is used, then $\kappa$ steps of RK($\tau$) are performed with a sampling period of $\tau$ each in order to predict the $\tau_u$-prediction step. Now MPC practice suggests that the precision of the prediction\footnote{To be distinguished from the precision of the simulation used in the certification and the tuning of the parameters which uses always the basic sampling period $\tau$.} might be degraded since in the closed-loop implementation, the state is periodically updated (each $\tau_u$ time units here), therefore, the closed-loop might be successful while  using $\texttt{n\_steps}$ ($\le \kappa$) inner steps of RK($\frac{\tau_u}{\texttt{n\_steps}}$) to get a step prediction over $\tau_u$. This suggests the following expression for the number of RK inner steps to yield a single $\tau_u$ step prediction\footnote{$\lceil r\rceil$ denotes the smallest integer greater than $r$.}:
\begin{equation}
\texttt{n\_steps}:=\lceil 1 + \mu_d(\kappa-1)\rceil
\end{equation} 
namely, $\mu_d=0$ yields a single RK($\tau_u$) \textit{large} step (low precision, faster computation) while $\mu_d=1$  yields $\kappa$ \textit{small} steps of RK($\tau$) (high precision, longer computation). This is illustrated in the figure below for the choices $\kappa=3$ and $\mu_d=0.5$ (hence leading to \texttt{n\_steps=2}):
\begin{center}
\begin{tikzpicture}
\draw[rounded corners, fill=Black!5, draw=white] (0.35,-0.6) rectangle (8.3,1.8);
\def\z{1}
\draw[->, >=latex] (0.5,0) -- node[below, pos=0.98]{\scriptsize \texttt{Time}} (8,0);
\draw[->, >=latex] (0.5,0) -- (0.5,1.4) node[right]{\footnotesize \color{RubineRed} Prediction's precision using \texttt{n\_steps=2}.};
\foreach \x in {1,...,14}
\filldraw (0.5*\x,0)  circle (1pt);
\draw[thin, MidnightBlue] (0.5,0) -- +(0,\z);
\draw[thin, MidnightBlue] (1,0) -- +(0,0.3*\z);
\draw[thin, MidnightBlue] (2,0) -- +(0,\z);
\draw[thin, MidnightBlue] (3.5,0) -- +(0,\z);
\draw[thin, MidnightBlue] (5,0) -- +(0,\z);
\draw[thin, MidnightBlue] (6.5,0) -- +(0,\z);
%-----
\draw[very thick, MidnightBlue] (0.5,0.8) -- (2,0.8);
\draw[very thick, MidnightBlue] (2,0.4) -- (3.5,0.4);
\draw[very thick, MidnightBlue] (3.5,0.6) -- (5,0.6);
\draw[very thick, MidnightBlue] (5,0.1) -- (6.5,0.1);
%-----
\draw[thin, RubineRed] (5.75, 0) -- (5.75,\z);
\draw[<->, RubineRed] (5,0.8*\z) -- node[above, RubineRed]{\footnotesize $\tau_p$} ++(0.75,0);
\draw[<->, RubineRed] (5.75,0.8*\z) -- node[above, RubineRed]{\footnotesize $\tau_p$} (6.5, 0.8*\z);
%-----
\draw[<->] (2,-0.2) -- node[MidnightBlue, midway, below]{\footnotesize $\tau_u=3\tau$}(3.5,-0.2);
\draw[<->, MidnightBlue] (0.5,0.2\z) -- node[midway, above]{\scriptsize $\tau$}(1,0.2\z);
%----
\end{tikzpicture}
\end{center} 

\vskip 1mm
\item[$\checkmark$] \hi{The prediction horizon's length} defined as a multiple of the \textbf{control updating period} $\tau_u=\kappa\tau$ through the parameter $N_\text{pred}$, namely the prediction horizon is given by: $$T = N_\text{pred}\times \kappa\times \tau=N_\text{pred}\times \tau_u$$ Notice that the prediction horizon's length is generally arbitrarily set by MPC practitioners unless one has in mind the standard stability conditions in which case, $T$ might be set to be greater than the not-easy-to-know \textit{reachability horizon}\footnote{The horizon length that enables to reach the implicitly targeted state from any initial state including in some subset of interest of the state space.}. Now not only this horizon is not known, moreover, one should remember that its use is only a sufficient condition for stability proof but that might lead to unnecessarily long prediction horizons that can threaten the real-time implementability. This is why this parameter is left to the algorithm which can find that shorter values might be needed to meet real-time implementability requirement. \vskip 1mm 
\item[$\checkmark$] \hi{The control horizon's length} $n_\text{contr}$. This is the number of updating instants over the prediction horizon before the control is frozen to the last value. In single shooting framework, this induces a decision variable of dimension $n_\text{contr}\times n_u$. This is a common practice for the definition of the control parameterization. Other options are obviously available\footnote{Such as representations via saturated functional basis or piece-wise linear interpolation via sub-sampling, etc. \cite{alamir2006stabilization}} that we skip here for the sake of clarity. The Figure below illustrates the used parameterization.\vskip 1mm
\begin{center}
\begin{tikzpicture}
\draw[fill=black!5, rounded corners, draw=white] (-0.2,2.1) rectangle (7.6,-0.6);
\draw[<->, >=latex] (0,1.8) node[right]{$\mathbf u(\cdot)$} -- (0,0) -- (7,0) node[below]{\scriptsize \texttt{time}};
\foreach \x in {1,...,8}{
\draw[fill=black] (0.75*\x,0) circle (1pt);
\draw[thin, dotted] (0.75*\x,0) -- +(0,1.5);
}
\draw[very thick, MidnightBlue] (0.75 * 0,0.4) -- (0.75 * 1,0.4);
\draw[very thick, MidnightBlue] (0.75 * 1,0.6) -- (0.75 * 2,0.6);
\draw[very thick, MidnightBlue] (0.75 * 2,0.95) -- (0.75 * 3,0.95);
\draw[very thick, MidnightBlue] (0.75 * 3,0.55) -- (0.75 * 8,0.55);
%----
\draw[<->, >=latex] (0.75 * 7,0.95) -- node[above]{\scriptsize $\tau_u$} (0.75 * 8,0.95);
\draw[fill=OrangeRed] (0*0.75, 0.4) circle (2pt);
\draw[fill=OrangeRed] (1*0.75, 0.6) circle (2pt);
\draw[fill=OrangeRed] (2*0.75, 0.95) circle (2pt);
\draw[fill=OrangeRed] (8*0.75, 0.55) circle (2pt);
%----
\draw[fill=OrangeRed] (0.75*2.5, 1.7) circle (2pt) node[right, RubineRed]{\footnotesize\  Degrees of freedom (\texttt{
n\_control=4)}};
\end{tikzpicture}
\end{center} 
\item[$\checkmark$] \hi{The weighting penalties $\rho_f$ and $\rho_\text{constr}$} used to enforce the terminal penalty and the soft constraints respectively. Notice that these two parameters which impact the \textit{stiffness} of the resulting optimization problems are almost never chosen in a rationale way. Instead, they are generally fixed to very high values that might be inspired by the understanding of their asymptotic\footnote{when the penalty goes to infinity.} impact on the stability and constraints satisfaction in a context where the real-time concern is absent. \vskip 1mm 
\item[$\checkmark$] \hi{The maximum number of iterations \texttt{max\_iter}} used in the optimization process. 
\end{itemize}
To summarize, the vector of NMPC design parameters that the algorithm is intended to tune is defined by:\vskip 1mm
\begin{center}
\begin{tikzpicture}
\node[rounded corners, fill=Gray!5, draw=Blue, inner ysep=5mm] (O){
\begin{minipage}{0.45\textwidth}
\color{Blue}
\begin{equation}
\pi := \begin{bmatrix}
\kappa\cr \mu_d \cr N_\text{pred} \cr  n_\text{contr} \cr  \rho_f \cr  \rho_\text{constr} \cr \texttt{max\_iter}
\end{bmatrix} \in \Pi:=[\underline\pi, \bar \pi]\subset \mathbb R^{n_\pi}\label{defdepi}
\end{equation} 	
\end{minipage} 
};
\node[rounded corners, fill=white, draw=MidnightBlue] at(O.north){\footnotesize The vector of design parameters};
\end{tikzpicture}
\end{center} 
\vskip 1mm 
\noindent in which  $\underline \pi$ and $\bar \pi$ are lower and upper bounds on the components of the design vector $\pi$ that are to be given as inputs to the tuning algorithm. \\ \ \\ This corresponds to a quite \textbf{rich and non convex} set of possibilities. For each candidate setting $\pi$, the corresponding NMPC has to be evaluated \textbf{for the targeted computation device and optimization algorithm} over a high number of relevant scenarios needed for the certification. Obviously, this cannot be done in an exhaustive way nor is it possible to consider an outer-loop optimizing $\pi$ while an inner-loop performs a probabilistic certification using a high number of scenarios. Some different heuristic should be derived which is explained in the remaining sections. 
\\ \ \\To summarize, the optimization problem that is addressed in the following section can be stated as follows:
\begin{center}
\begin{tikzpicture}
\node[rounded corners, fill=Black!5, inner xsep=2mm, inner ysep=4mm, draw=Blue](O){\color{Blue}
\begin{minipage}{0.45\textwidth}
\textbf{Given} the following items:\vskip 1mm
\begin{itemize}
\item The problem's ingredients described in Section \ref{sec_inputalgo};\vskip 1mm
\item An admissible set $\Pi:=[\underline\pi, \bar \pi]$;\vskip 1mm
\item A computation target;
\item A given optimization algorithm,\vskip 1mm
\end{itemize}    
\textbf{Derive} a \textit{\color{RubineRed} tractable heuristic} that:
\begin{itemize}
\item \underline{Either} yields a rational choice of the vector of parameters $\pi$ addressing \textbf{stability, constraints satisfaction and real-time implementability} concerns;\vskip 1mm
\item \underline{Or} it suggests that these requirements cannot be met given the data of the problem. 
\end{itemize} 
  
\end{minipage} 
};
\node[rounded corners, fill=white, draw=MidnightBlue] at(O.north){\scriptsize Problem Statement};
\end{tikzpicture}
\end{center}
Such a solution is proposed in the following section. 
\section{The proposed tuning algorithm}\label{sec_theAlgo}
In this section, the parameterization of the set of available degrees of freedom contained in \eqref{defdepi} is first explained in Section \ref{subParam} in ordre to derive a tractable tuning algorithm. Then the overall computation architecture is sketched in Sections \ref{suboverall} and Section \ref{subselectioncrit} leading to an ideal formulation that is then relaxed in Section \ref{sub_optimal_tractable} to yield the finally proposed tractable algorithm.
\subsection{Parameterization of the set of NMPC design parameters} \label{subParam}
\begin{figure}[H]
\begin{center}
\begin{tikzpicture}
\def\x{0}
\def\r{4}
\def\w{3}
\def\h{1.8}
%----
\draw[fill=Black!5, rounded corners, draw=white] (-0.3,\h+0.3) rectangle (\r+\w+0.2, -0.6);
%----
\draw[<->, >=latex] (\x,\h) node[right]{\footnotesize \texttt{cpu}} -- (\x,0) -- (\x+\w,0) node[below, pos=0.8]{\footnotesize \texttt{quality}};
\draw[MidnightBlue, thick, postaction={decorate, decoration={markings, mark=at position 0.7 with {\arrow{>}}}}]  (\x,0.1)  to[bend right] node[right, pos=0.7]{\scriptsize $\pi_i\uparrow$} (\x+0.7*\w,0.9*\h);
\draw[MidnightBlue, thick, postaction={decorate, decoration={markings, mark=at position 0.3 with {\arrow{>}}}}]  (\x,0.2) to[bend left] node[right, pos=0.3]{\scriptsize $\pi_i\uparrow$} (\x+0.6*\w,0.95*\h);
\node[MidnightBlue] at (\x+1.8,0.3){\scriptsize \underline{\textsc{Type(+)}}};
%----
\draw[<->, >=latex] (\r,\h) node[right]{\footnotesize \texttt{cpu}} -- (\r,0) -- (\r+\w,0) node[below, pos=0.8]{\footnotesize \texttt{quality}};
\draw[RubineRed, thick, postaction={decorate, decoration={markings, mark=at position 0.7 with {\arrow{>}}}}]  (\r,0.1)  to[bend right] node[right, pos=0.7]{\scriptsize $\pi_i\downarrow$} (\r+0.7*\w,0.9*\h);
\draw[RubineRed, thick, postaction={decorate, decoration={markings, mark=at position 0.3 with {\arrow{>}}}}]  (\r,0.2) to[bend left] node[right, pos=0.3]{\scriptsize $\pi_i\downarrow$} (\r+0.6*\w,0.95*\h);
\node[RubineRed] at (\r+1.8,0.3){\scriptsize \underline{\textsc{Type(--)}}};
\end{tikzpicture}
\end{center} 
\caption{\color{Blue} Any component $\pi_i$ of the design vector $\pi$ is either of type(+) [left] or of type (--) [right].}\label{lescurves}
\end{figure}
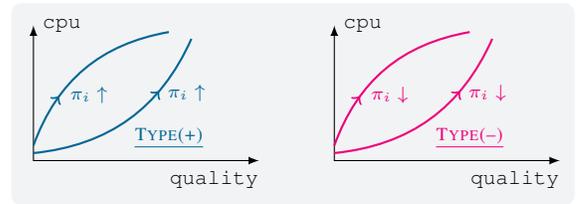

\noindent In order to break the complexity of the search over the domain $\Pi$ defined in \eqref{defdepi}, a first observation is worth making that can be stated as follows: \\ \ \\
\begin{minipage}{0.01\textwidth}
\color{Gray}\rule{1mm}{23mm}	
\end{minipage} 
\begin{minipage}{0.47\textwidth}
\color{Black!80}
\textbf{\underline{Two types of components:}}\vskip 1mm 
Each component $\pi_i$ of $\pi$ is of one of the two types shown in Figure \ref{lescurves}, namely, either the pairs (\texttt{cpu/quality}) are increasing functions of $\pi_i$ (\textsc{Type(+)}) or they are decreasing functions of $\pi_i$ (\textsc{Type(--)}). \vskip 1mm \hrule 
\end{minipage} 
\noindent Given the definition \eqref{defdepi} of $\pi$, it can be easily checked that all the components are of type(+) except $\pi_1=\kappa$ which is of type (--).\\ \ \\
%\begin{table}[H]
%  \begin{center}
%    \caption{Types of the $\pi$ vector's components}
%    \label{tabletype}
%    \begin{tabular}{|l|c|l|c|} % <-- Alignments: 1st column left, 2nd middle and 3rd right, with vertical lines in between
%    \hline 
%      \textbf{Components} & \textbf{Type} & \textbf{component} & \textbf{Type}\\
%      \hline
%      $\kappa$ & (--) & $\rho_f$ & (+)\\
%      $\mu_d$ & (+) & $\rho_\text{constr}$ & (+)\\
%      $N_\text{pred}$ & (+) & \texttt{max\_iter}& (+)\\
%      $n_\text{contr}$ & (+)  &  & \\
%      \hline
%    \end{tabular}
%  \end{center}
%\end{table}
\noindent Notice however that for given a component, the shape of the monotonicity mentioned above for a given NMPC problem is difficult to know a priori. That is the reason why the shapes of the curves shown in Figure \ref{lescurves} are determined hereafter via a set of random sampling of a \hi{shaping parameter vector}\footnote{The appropriate shape may obviously not be the same for all the NMPC design parameters (components of $\pi$)}:
\begin{equation*}
\mathbf{\sigma}:=[\sigma_1,\dots,\sigma_{n_\pi}]\in \mathcal S^{n_\pi}\end{equation*}
were $n_\pi:=\text{card}(\pi)(=7$ in the current setting) while $\mathcal S$ is a set of allowed integers bounded by $\bar\sigma$ of the form:
\begin{equation*}
\mathcal S:=\{-\bar \sigma,\dots, -1, +1, \dots \bar\sigma\}
\end{equation*} 
For each sampled choice of the shaping parameter vector $\mathbf{\sigma}$, the shape of the curve representing the monotonicity of $\pi_i$ is defined via a function $\phi_{\sigma_i}(\alpha):[0,1]\rightarrow [0,1]$ which depends on a a scalar parameter $\alpha\in [0,1]$: 
\begin{equation}
\phi_{\sigma_i}(\alpha):=
\left\{
\begin{array}{ll}
\alpha^{\sigma_i} & \text{if $\sigma_i >0\quad$ \footnotesize (positive curvature)} \\
\alpha^{\frac{1}{\vert \sigma_i\vert}} & \text{if $\sigma_i <0\quad$ \footnotesize (negative curvature)} 
\end{array}
\right.
\end{equation} 
and this curve is used to define the value of $\pi_i$ according to:
\begin{equation}
\pi_i(\alpha):=\left\{
\begin{array}{ll}
(1-\phi_{\sigma_i}(\alpha))\underline{\pi}_i+\phi_{\sigma_i}(\alpha)(\bar \pi_i-\underline\pi_i) &\text{\scriptsize Type (+)} \tiny\\
(1-\phi_{\sigma_i}(\alpha))\bar{\pi}_i+\phi_{\sigma_i}(\alpha)(\underline\pi_i-\bar \pi_i) &\text{\scriptsize Type (--)} 	
\end{array}
\right.\label{defdepialphasigma}
\end{equation}  
Consequently, the vector of shaping parameters $\mathbf{\sigma}$ is to be sampled in the set $\mathcal S^{n_\pi}$.\\ \ \\
Figure \ref{fig_shapes} illustrates the above definitions for four different configurations of the pair ($\sigma$, type) showing obviously that the sign of $\sigma_i$ determines the curvature of the increasing (type+) or decreasing (type--) allure of the $\pi_i$ as a function of the design parameter $\alpha$:
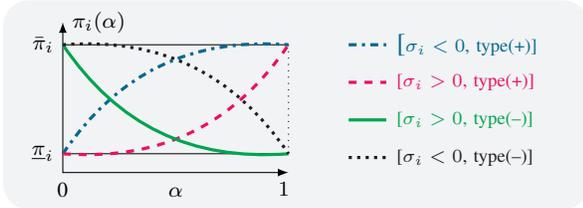
\begin{figure}[H]
\begin{center}
\begin{tikzpicture}
%---
\draw[fill=Black!5, draw=white, rounded corners=10pt] (-0.8, 2.3) rectangle (7,-0.5);
%---
\draw[<->, >=latex] (0,2) node[right]{\footnotesize $\pi_i(\alpha)$} -- (0,0) node[below]{\footnotesize 0} -- node[below=2pt, pos=0.5]{\footnotesize $\alpha$} (3,0) node[below, pos=0.98]{\footnotesize $1$};
\draw[very thin] (0, 0.25) coordinate (O1) node[left]{\footnotesize $\underline \pi_i$} -- (3,0.25) coordinate (O4);
\draw[very thin] (0, 1.7) coordinate (O3) node[left]{\footnotesize $\bar \pi_i$} -- (3,1.7) coordinate (O2);
\draw[dotted] (3,0.25) -- (3, 1.7);
%--- draw the curve
\draw[very thick, MidnightBlue, dash dot] (O1) to[bend left] (O2);
\draw[very thick, OrangeRed, dashed] (O1) to[bend right] (O2);
\draw[very thick, Green] (O3) to[bend right] (O4);
\draw[very thick, Black, dotted] (O3) to[bend left] (O4);
%---- draw the legend 
\draw[very thick, MidnightBlue, dash dot] (3.8, 1.7) -- ++(0.5, 0) node[right] {[\scriptsize $\sigma_i<0$, type(+)]};
\draw[very thick, OrangeRed, dashed] (3.8, 1.2) -- ++(0.5, 0) node[right] {\scriptsize [$\sigma_i>0$, type(+)]};
\draw[very thick, Green] (3.8, 0.7) -- ++(0.5, 0) node[right] {\scriptsize [$\sigma_i>0$, type(--)]};
\draw[very thick, Black, dotted] (3.8, 0.2) -- ++(0.5, 0) node[right] {\scriptsize [$\sigma_i<0$, type(--)]};
\end{tikzpicture}
\end{center} 
\caption{\color{Blue} Typical shapes of $\pi_i(\alpha)$ depending on the sign of $\sigma_i$ and the type of the component $\pi_i$.}\label{fig_shapes}
\end{figure}
\noindent This figure only shows qualitative examples of the available shapes, by modifying the sampled $\mathbf \sigma$, one can vary the stiffness of the represented curves. \\ \ \\
\noindent It is worth noticing that the parameterization defined by \eqref{defdepialphasigma} is such that, \textbf{regardless of the type of the component} being considered, \hi{low values of $\alpha\in [0,1]$ correspond to low computational complexity and low performance levels} while high values induce high computation times and better performance levels should the latter computations be possible within the updating period. This property plays a crucial rule in the forthcoming computation since this monotonicity enables dichotomy-based iterations to be used. \\ \ \\
It is possible to summarize the previous discussion as follows:\\ \ \\ 
\begin{minipage}{0.01\textwidth}
\color{Gray}\rule{1mm}{36mm}	
\end{minipage} 
\begin{minipage}{0.47\textwidth}
\color{Black!80}
\textbf{\underline{$\alpha$-parameterized spanning the space of NMPC settings:}}\vskip 1mm
Each paire $(\mathbf \sigma, \alpha)\in \mathcal S^{n_\pi}\times [0,1]$ defines a specific NMPC setting since all the design components of $\pi$ defined by \eqref{defdepi} are fixed by \eqref{defdepialphasigma}. Moreover, when $\alpha$ spans $[0,1]$ the components of $\pi$ travel from one extreme value to the other in the sense of increasing computational complexity and ideally increasing quality should the concern regarding the real-time implementability be discarded.\vskip1mm \hrule
\end{minipage} 
\vskip 3mm 
\subsection{The ideal computation architecture at a glance}\label{suboverall}
\noindent The remaining task is to find the \textit{best} $(\mathbf\sigma,\alpha)\in  \mathcal S^{n_\pi}\times [0,1]$ in terms of the control objective and implementability. Obviously, the two components of the design, namely $\sigma$ and $\alpha$ are to be handled differently. Indeed, it seems reasonably easy to select the scalar $\alpha$ for a given $\sigma$ as the underlying \textit{indicators} vary monotonically in $\alpha$. On the contrary, it is unrealistic to attempt a complete and rigorous optimization of $\sigma\in  \mathcal S^{n_\pi}$ because of the highly combinatorics and the involved computational burden for each candidate value.
\begin{figure}[H]
\begin{center}
\begin{tikzpicture}
\draw[rounded corners=10pt, fill=Black!5, anchor=north west, draw=white] (-3.2,2) rectangle (5.6,-3);
\def\r{0.5}
\node[draw=Blue, rectangle, rounded corners, anchor=north west, minimum width=3.9cm, minimum height=1.6cm, name=rect, very thick,, fill=white, opacity=0.9] at(-0.4, 1.5*\r){};
\draw[thick,postaction={decorate,decoration={markings,
            mark=at position 0.75 with {\arrow{stealth}}
          }
        }]
        (-0.25,0) arc (180:-90:\r);
\node at(0.5*\r,0)(O){$\alpha$};
\node[anchor=west] at(2*\r-0.25, 0){\footnotesize  Compute $\hat\alpha(\sigma^{(j)}\vert\mathcal A)$}; 
\node[anchor=west] at(2*\r, -0.4){\scriptsize  (Section \ref{subselectioncrit})};       
\draw[->,>=latex, rounded corners, RubineRed, very thick] (rect.south) -- ++(0,-0.4) -- node[below, RubineRed]{\footnotesize\texttt{N\_trials}} ++(-3.3,0) -- node[fill=Black!5]{\footnotesize Sample $\sigma^{(j)}\in \mathcal S^{n_\pi}$} ++(0,2.5) -| node[RubineRed, above, pos=.3]{\footnotesize $\sigma^{(j)}$} (rect.north);
\draw[<-, very thick, >=latex, rounded corners, MidnightBlue] (rect.60) -- ++(0,0.6) -- node[above, MidnightBlue]{\footnotesize Set of scenarios $\mathcal A$} ++(2.2,0);
%--
\draw[->,>=latex, very thick] (rect.290) -- node[right]{\scriptsize Best cost $\vert \sigma^{(j)}$} ++(0,-1) node[below, draw, thin, rounded corners, fill=white](J){\footnotesize $\Bigl\{\mathcal J(\sigma^{(j)}\vert \mathcal A)\Bigr\}_{j=1}^\texttt{N\_trials}$};
\draw[->,>=latex] (J.east) -- node[below]{\scriptsize $\min_{j}$} node[above]{\scriptsize \eqref{defdesigmastar}-\eqref{defdealphastar}} +(1.1,0) node[right]{\footnotesize $(\sigma^\star,\alpha^\star)$};
\draw[<-,>=latex,thick,MidnightBlue] (rect.10) -- ++(0.5,0) node[right]{\scriptsize $(m,\gamma,\varepsilon)$};
\draw[<-,>=latex,thick,MidnightBlue] (rect.0) -- ++(0.5,0) node[right]{\scriptsize\texttt{dev\_acc}};
\draw[<-,>=latex,thick,MidnightBlue] (rect.350) -- ++(0.5,0) node[right]{\scriptsize\texttt{c\_max}};
\end{tikzpicture}
\end{center} 	
\caption{\color{Blue} Architecture of the two-layer algorithm for the determination of the NMPC setting paire $(\sigma^\star, \alpha^\star)$ for a given set of representative scenarios $\mathcal A$.}\label{fig_at_a_glance}
\end{figure}
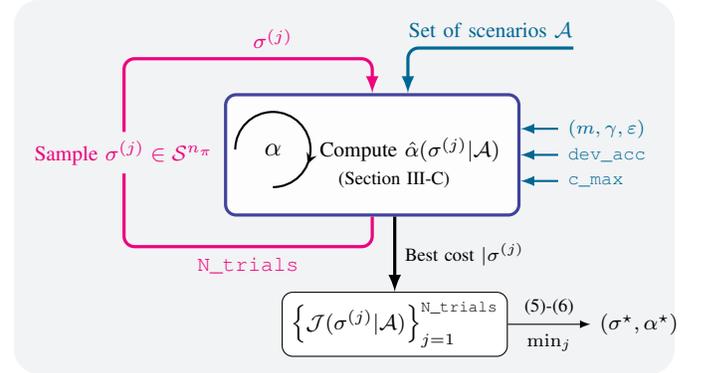
\noindent That is the reason why, the two-layer architecture shown in Figure \ref{fig_at_a_glance} is adopted in which \texttt{N\_trials} values of the shaping parameter vectors $\bigl\{\sigma^{(j)}\in \mathcal S^{n_\pi}\bigr\}_{j=1}^\texttt{N\_trials}$ are sampled in $\mathcal S^{n_\pi}$ and for each sampled $\sigma^{(j)}$, a scalar constrained optimization on $\alpha\in [0,1]$ is performed to check whether there is at least one value that meets the requirements (over a pre-defined set $\mathcal A$ of scenarios as explained later on) and if any, find the \textit{optimal} one, denoted by $\hat\alpha(\sigma^{(j)}\vert \mathcal A)$. These requirements and the associated scalar optimization problem are explained in Section \ref{subselectioncrit}. The computation yields the best cost $\mathcal J(\sigma^{(j)}\vert \mathcal A)$ given $\sigma^{(j)}$. The \textit{best} NMPC design setting paire $(\sigma^\star, \alpha^\star)$ is therefore obtained by:
\begin{align}
\sigma^\star &\leftarrow \text{arg}\min\Bigl\{\mathcal J(\sigma^{(j)}\vert\mathcal A)\quad \footnotesize j\in \{1,\dots,\texttt{N\_trials}\}\Bigr\} \label{defdesigmastar}\\
\alpha^\star &\leftarrow \hat\alpha(\sigma^\star\vert \mathcal A)\label{defdealphastar}
\end{align}  
This principle is depicted in Figure \ref{fig_at_a_glance} while the details of the boxes content is explained  in the following sections\footnote{In particular, the significance of $m,\gamma, \varepsilon$, \texttt{dev\_acc} and \texttt{c\_max} is given in Section \ref{subselectioncrit}.}. Let us first focus on the inner optimization loop that computes $\hat\alpha(\sigma^{(j)}\vert \mathcal A)$. This is the aim of the following section. 
\subsection{The admissibility criteria: Computing $\hat\alpha(\sigma\vert \mathcal A)$}\label{subselectioncrit}
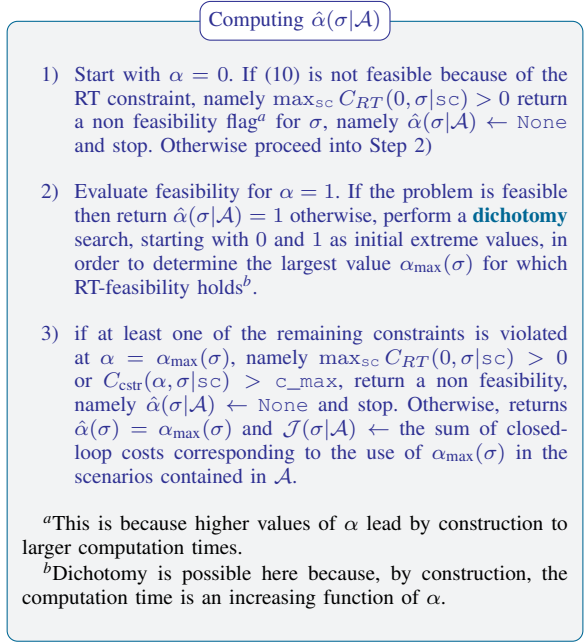
\begin{figure}[h]
\begin{center}
\begin{tikzpicture}
\node[rounded corners, fill=Black!5, draw=MidnightBlue, inner xsep=2mm, inner ysep=4mm]at(0,0)(O){
\begin{minipage}{0.4\textwidth}
\footnotesize \color{Blue}
\vskip 2mm
\begin{itemize}
 \item[1)] Start with $\alpha=0$. If \eqref{prob} is not feasible because of the RT constraint, namely $\max_\texttt{sc} C_{RT}(0,\sigma\vert \texttt{sc})>0$ return a non feasibility flag\footnote{This is because higher values of $\alpha$ lead by construction to larger computation times.} for $\sigma$, namely $\hat\alpha(\sigma\vert \mathcal A)\leftarrow \texttt{None}$ and stop. Otherwise proceed into Step 2)\vskip 3mm
\item[2)] Evaluate feasibility for $\alpha=1$. If the problem is feasible then return $\hat\alpha(\sigma\vert \mathcal A)=1$ otherwise, perform a \hi{dichotomy} search, starting with $0$ and $1$ as initial extreme values, in order to determine the largest value $\alpha_\text{max}(\sigma)$ for which RT-feasibility holds\footnote{Dichotomy is possible here because, by construction, the computation time is an increasing function of $\alpha$.}. \vskip 3mm
\item[3)] if at least one of the remaining constraints is violated at $\alpha=\alpha_\text{max}(\sigma)$, namely $\max_\texttt{sc} C_{RT}(0,\sigma\vert \texttt{sc})>0$ or $C_\text{cstr}(\alpha,\sigma\vert \texttt{sc})> \texttt{c\_max}$, return a non feasibility, namely $\hat\alpha(\sigma\vert \mathcal A)\leftarrow \texttt{None}$ and stop. Otherwise, returns $\hat\alpha(\sigma)=\alpha_\text{max}(\sigma)$ and $\mathcal J(\sigma\vert \mathcal A)\leftarrow$ the sum of closed-loop costs corresponding to the use of $\alpha_\text{max}(\sigma)$ in the scenarios contained in $\mathcal A$.\vskip 3mm
\end{itemize}

\end{minipage} 
};
\node[fill=white, draw=Blue, rounded corners] at(O.north){\color{Blue} \footnotesize Computing $\hat\alpha(\sigma\vert \mathcal A)$};
\end{tikzpicture}
\end{center} 
\caption{\color{Blue} The algorithm exploiting the specificity of the constrained scalar optimization problem \eqref{prob} to compute the optimal $\hat\alpha(\sigma\vert \mathcal A)$ for a given shaping parameter vector $\sigma$ and a set of scenarios $\mathcal A$.}\label{fig_algo_alphahat}
\end{figure}

\noindent For a given shaping vector candidate value $\sigma$, admissible values of $\alpha$ are those that makes the NMPC setting defined by $(\sigma, \alpha)$ compatible with the requirements regarding the real-time implementation, the  constraints satisfaction and the stability. This section gives a concrete quantification for these criteria. Before doing so, it is worth underlying that the assessment of a candidate choice $(\sigma,\alpha)$ is obtained by examining numerical experiments consisting each of a finite number of  $m$ successive closed-loop steps simulated for a given  scenario \texttt{sc} in a set of scenarios $\mathcal A$. \\ \ \\ More precisely, \hi{Considering $m$ updating steps} over a closed-loop simulation scenario \texttt{sc} $\in \mathcal A$ corresponding to a given initial state, a given model's parameters $p$ and an exogenous task related parameter vector $q$ (see Section \ref{sec_inputalgo}), there are three concerns when it comes to evaluate the successful (or not) use of the NMPC setting on this specific scenario, namely:
\begin{itemize}
\item[$\checkmark$] \hi{The real-time feasibility} which can be stated by the following constraint\footnote{$\lfloor \xi\rfloor_+:=\max(0,\xi)$}: 
\begin{align}
&\text{\color{Blue} RT-feasibility} \label{req1} \\ &C_{RT}(\alpha,\sigma\vert \texttt{sc}):=\max_{k=1,\dots,m}\left\lfloor \dfrac{\tau_\text{solver}}{\tau_u}(k)-1\right\rfloor_+ =0 \nonumber
\end{align} 
where $\tau_\text{solver}(k)$ denotes the time needed to solve the underlying $k$-th optimization problem encountered in the scenario and given the chosen maximum number of iterations $\pi_7(\alpha)=\phi_{\sigma_7}(\alpha)$ corresponding to the currently evaluated $\sigma$. Indeed, if the above expression is equal to $0$, this means that for all $k$, one has $\tau_\text{solver}(k)\le \tau_u$. \\ \ \\ Notice that be replacing $\tau_u$ in \eqref{req1} by $\texttt{dev\_acc}\times \tau_u$, real-time implementatbility can be checked for a targeted computational resource that is \texttt{dev\_acc} faster than the one on which the evaluation is done (or \texttt{dev\_acc} times slower for \texttt{dev\_acc} $< 1$). \vskip 1mm
\item[$\checkmark$] \hi{The contraction property} that implicitly assumes that the MPC formulation is such that a decrease of the cost function is expected over the closed-loop trajectory although not necessarily at each step (as in contraction-based formulation for instance \cite{alamir2017contraction}). Therefore, using $m$-step contraction horizon, the associated requirement writes:
\begin{align}
&\text{\color{Blue} $\gamma$-Contraction}\label{req2} \\ &C_\gamma(\alpha,\sigma\vert \texttt{sc}):=\lfloor J_{ol}(m)-\gamma J_{ol}(1)\rfloor_+=0\nonumber 
\end{align} 
where $J_{ol}(k)$ is the \textit{best open-loop cost value} returned by the solver at the $k$-th updating computation\footnote{Given the allowed maximum number of iterations corresponding to the design parameter $\pi_7=\texttt{max\_iter}$ [see \eqref{defdepi}].} while $\gamma\in (0,1)$ is some predefined contraction rate.\vskip 1mm 
\item[$\checkmark$] \hi{The constraints satisfaction} which applies mainly to soft exact penalty constraints since it is reasonably assumed that the input hard constraints are structurally enforced by the optimization algorithm. Therefore, the constraints satisfaction assessment criterion becomes:
\begin{align}
&\text{\color{Blue} Constraints satisfaction}\label{req3} \\ 
&C_\text{cstr}(\alpha,\sigma\vert \texttt{sc}):=\max_{k=1,\dots,m}\lfloor \max_{i\le n_c} c_i(k)\rfloor_+ = 0\nonumber
\end{align} 
\end{itemize} 
Notice that the above mentioned \textit{success conditions}, namely \eqref{req1}, \eqref{req2} and \eqref{req3}, concern a specific scenario \texttt{sc}.  Now obviously, the overall assessment of a candidate NMPC setting paire  $(\sigma,\alpha)\in \mathcal S^{n_\pi}\times [0,1]$ has to be scenario-independent. This is the reason why a set $\mathcal A$ of \textit{representative scenarios} is considered. As it is discussed later, the cardinality of this set can be determined following the probabilistic certification formulas\footnote{These formulas give the number of scenarios to use in order to state with a given confidence level that the probability of violating the requirements is lower than a predefined level.} \cite{alamo2009}. \\ \ \\ For the remainder of the presentation, it is hence assumed that one disposes of a generator of $\mathcal A\leftarrow \texttt{Generate\_A}(n_\text{sc})$ that admits the cardinality of the required output set as argument. The set $\mathcal A$ is used to compute the \textit{best} value $\hat\alpha(\sigma\vert \mathcal A)$ \hi{for a given $\sigma$} by solving the following constrained scalar optimization problem:
\begin{center}
\begin{tikzpicture}
\node[rounded corners, fill=Gray!10, inner xsep=2mm, inner ysep=3mm](O){
\begin{minipage}{0.48\textwidth}
\small
\begin{subequations}\label{prob}
\begin{align}
\hat\alpha(\sigma\vert \mathcal A)\leftarrow \  &{\color{MidnightBlue} \max_{\alpha\in [0,1]}}\left[\alpha\right]\label{op1}\\
&\text{\color{MidnightBlue} under} \left\vert 
\begin{array}{l}
\displaystyle{\max_{\texttt{sc}\in \mathcal A}}\ C_{RT}(\alpha,\sigma\vert \texttt{sc})=0\cr 	
\displaystyle{\max_{\texttt{sc}\in \mathcal A}}\ C_{\gamma}(\alpha,\sigma\vert \texttt{sc})=0\cr 
\displaystyle{\max_{\texttt{sc}\in \mathcal A}}\ C_\text{cstr}(\alpha,\sigma\vert \texttt{sc})\le \texttt{c\_max}
\end{array}
\right.\label{op1_cstr}
\end{align} 	
\end{subequations}
\end{minipage} 
};	
\node[rounded corners, fill=white, draw=Blue] at(O.north){\footnotesize$\mathcal P(\sigma, \mathcal A)$};
\end{tikzpicture}
\end{center} 
where \texttt{c\_max} is a threshold on the level of possible violation of the the soft constraints. Notice that the rationale in maximizing $\alpha$ relies on the very definition of the parameterization described in Section \ref{subParam} which leads to better open-loop solutions for for higher values of $\alpha$.\\ \ \\ Although one can view \eqref{prob} as a general constrained optimization problem, the specificity of the problem enables a simple method to derive quite good suboptimal solutions. This is even mandatory given the computation effort needed to evaluate the terms involved in \eqref{prob} for any candidate value of $\alpha$ (since all the scenarios included in $\mathcal A$ are involved). The specific process which is mainly based on a dichotomic search is described in details by the algorithm of Figure \ref{fig_algo_alphahat}.
%------------------------
%------------------------
\ \\
\noindent This algorithm delivers for each candidate shaping parameter vector $\sigma$ and for a given set of scenarios $\mathcal A$, an associated $\hat\alpha(\sigma\vert \mathcal A)$ together with the associated optimal cost $\mathcal J(\sigma\vert \mathcal A)$. \ \\ \ \\ As already shown in Figure \ref{fig_at_a_glance}, the final solution $(\sigma^\star(\mathcal A),\alpha^\star(\mathcal A))$ is finally determined by \eqref{defdesigmastar}-\eqref{defdealphastar}, namely taking the \textit{best} shaping parameter vector $\sigma^\star$ among those obtained through the randomly sampled \texttt{N\_trials} values $\sigma^{(j)}$ for $j=1,\dots, \texttt{N\_trials}$ together with the corresponding computed $\alpha^\star:=\hat\alpha(\sigma^\star\vert \mathcal A)$.\\ \ \\ Unfortunately, this process is computationally very expensive as explained in the next section where an alternative sub-optimal but far more tractable alternative is proposed. 
\subsection{A sub-optimal tractable algorithm}\label{sub_optimal_tractable}
\noindent The previous section describes an ideal formulation for the computation of the optimal NMPC setting parameters $(\sigma^\star, \alpha^\star)$. Unfortunately, in its present form, this formulation requires a huge amount of computation. Indeed for each $\sigma$ a dichotomic search has to be performed to compute the optimal $\hat\alpha(\sigma\vert \mathcal A)$ (if any) in which the expressions involved in the constraints \eqref{op1_cstr} has to be evaluated through $m$-steps closed-loop simulations and this,  \hi{for all the scenarios included in the high cardinality set $\mathcal A$}. Assuming a precision of $\epsilon$ on $\alpha$, this induces a \textit{worst case} number of optimal control problems that compares to 
\begin{equation}
\texttt{N\_trials} \times \text{card}(\mathcal A)\times m \times \log(1/\varepsilon)
\end{equation}  
Note that $\text{card}(\mathcal A)$ is determined by the precision ($\eta\in (0,1)$) and the confidence ($\delta\in (0,1)$) parameters required for the certification of the success when using the corresponding setting\footnote{More precisely, assuming a property $P$ to be certified, the certification statement would be: \textit{it can be guaranteed with a probability $1-\delta$ (confidence in the statement) that the probability of having $P$ satisfied is greater than $1-\eta$ (precision of the property satisfaction).}} (see table \ref{lesNtab}). 
\begin{table}[H]
\begin{center}
\begin{tabular}{lcccc} \hline
    {\texttt{N\_trials}} & {$\eta=0.1$} & {$\eta=0.05$} & {$\eta=0.01$} & {$\eta=0.001$} \\ \hline
    1  & 132 &       264 &      1317 &       13164 \\
    5  & 154 &       308  &     1536    &    15354\\
    10  & 163 &        326 &         1628    &    16280\\
    100  & 193 &       386 &       1930  &      19299\\
    1000  & 223 &       445 &       2225 &       22249\\
 \hline
\end{tabular}
\end{center} 
\ \\
\caption{\color{Blue} The required card($\mathcal A$) as a function of the precision parameter $\eta$ for a confidence parameter of $\delta=10^{-3}$ and a number of admitted failure $=1$.} \label{lesNtab} 
\end{table}
\ \\ 
That is the reason why a sub-optimal version of the formulation is adopted in which, the process is split into two sub-processes, namely: \\
\begin{itemize}
\item[$\checkmark$] First, a randomly generated set $\mathcal A_0$ containing a reduced number of scenarios, namely $n_0:=$ card($\mathcal A_0$) $\ll$ card($\mathcal A$) is considered for which the previously described optimal formulation is applied in order to determine $\hat\alpha(\sigma^{(j)}\vert \mathcal A_0)$ for $j=1,\dots,\texttt{N\_trials}$.\vskip 2mm
\item[$\checkmark$] \hi{The so obtained $\hat\alpha(\sigma^{(j)}\vert \mathcal A_0)$ are now frozen} and the $\sigma^{(j)}$ are individually checked\footnote{Which means that we can use the first line of Table \ref{lesNtab}} over a sequence of subsets $\mathcal A^{[\ell]}$ that forms a partition of the original set $\mathcal A$, namely: $$\mathcal A:=\bigcup_{\ell=1}^{n_s}\mathcal A^{[\ell]}$$
and {\color{RubineRed} the associated optimal costs are summed up} over the subsets $\mathcal A ^{[\ell]}$ in order to ultimately form the corresponding cost $\mathcal J(\sigma^{[j]}\vert \mathcal A)$. \\ \ \\ \hi{During this process, as soon as a failure is detected for some $\sigma^{[j]}$ for one of the scenarios of a subset $\mathcal A^{[\ell]}$, this  $\sigma^{[j]}$ is removed from the set of candidate values and is no more considered for the remaining subsets} $\mathcal A^{[\ell+\cdot]}$.   
\end{itemize} 
\ \\
This solution hugely reduces the number of evaluations since dichotomy search is restricted to $\mathcal A_0$ on one hand and the number uselessly visited evaluations is drastically reduced by progressively removing unsuccessful $\sigma^{[j]}$ after failure on intermediate low cardinality $\mathcal A^{[\ell]}$ on the other hand. Obviously the counter-part of this simplification is that the frozen values of $\hat\alpha(\sigma^{(j)})$ do not take into account all the scenarios of $\mathcal A$ and might hence be wrongly biased by the small number $n_0$ of initial scenarios contained in $\mathcal A_0$.
\ \\ \ \\
\noindent This achieves the presentation of the design algorithm. In the next section, the proposed \texttt{python}-based package is briefly described.
\section{Brief Description of the \texttt{python}-package}\label{sec_package}
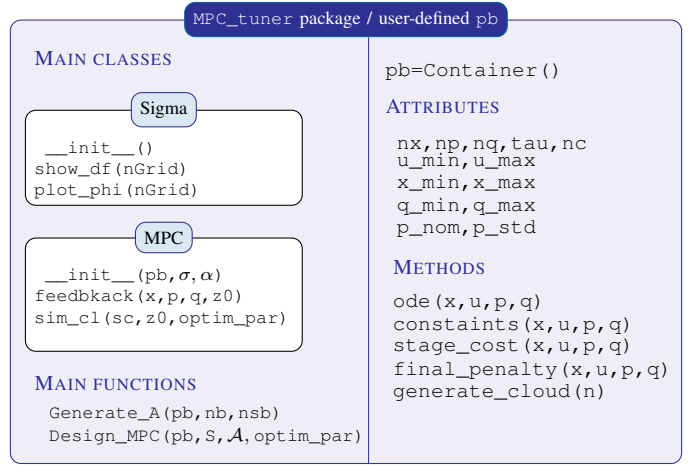
\begin{figure}
\begin{center}
\begin{tikzpicture}
\node[rounded corners, fill=Blue!5, inner sep=2mm, anchor=north west, draw=Blue, text width=0.47\textwidth, text height=5.5cm] at(0,0)(O){};
\node[rounded corners, draw=MidnightBlue, fill=Blue] at(O.north)(T){\color{white}\scriptsize \texttt{MPC\_tuner} package / user-defined \texttt{pb}};
%-----------------
\node[anchor=north west] at($(O.north west)+(0.2,-0.3)$){\footnotesize \sc \color{Blue} Main classes};
%----- class Sigma
\node[rounded corners, draw=black, fill=white, anchor=north west, inner ysep=1mm] at($(O.north west)+(0.2,-1.2)$) (S){
\begin{minipage}{0.19\textwidth}
\vskip 3mm
\scriptsize \texttt{
\_\_init\_\_()\\
show\_df(nGrid)\\
plot\_phi(nGrid)
}
\end{minipage} 
};
\node[fill=MidnightBlue!10, rounded corners, draw=Blue] at(S.north){\scriptsize Sigma};
%----- class MPC 
\node[rounded corners, draw=black, fill=white, anchor=north west, inner ysep=1mm] at($(O.north west)+(0.2,-2.9)$) (M){
\begin{minipage}{0.19\textwidth}
\vskip 3mm
\scriptsize \texttt{
\_\_init\_\_(pb,$\sigma, \alpha$)\\
feedbkack(x,p,q,z0)\\
sim\_cl(sc,z0,optim\_par)
}
\end{minipage} 
};
\node[fill=MidnightBlue!10, rounded corners, draw=Blue] at(M.north){\scriptsize MPC};
%------- The global functions
\node[anchor=north west] at($(M.south west)+(0,-0.2)$){\footnotesize \sc \color{Blue} Main functions};
\node[anchor=north west] at($(M.south west)+(0.2,-0.6)$){\scriptsize \texttt{Generate\_A(pb,nb,nsb)}};
\node[anchor=north west] at($(M.south west)+(0.2,-0.9)$){\scriptsize \texttt{Design\_MPC(pb,S,$\mathcal A,\texttt{optim\_par}$)}};
%------ vertical line 
\draw[very thin, Blue] ($(T.south)+(0.3,0)$) coordinate (U) -- ($(O.south)+(0.3,0)$);
%------ User defined object 
\node[anchor=north west] at($(U.south west)+(0.1,-0.2)$){\footnotesize  \texttt{pb=Container()}};
\node[anchor=north west] at($(U.south west)+(0.1,-0.7)$) (A) {\footnotesize  \color{Blue} \textsc{Attributes }};
\node[anchor=north west] at($(U.south west)+(0.25,-1.2)$){\footnotesize  \texttt{nx,np,nq,tau,nc}};
\node[anchor=north west] at($(U.south west)+(0.25,-1.4)$){\footnotesize  \texttt{u\_min,u\_max}};
\node[anchor=north west] at($(U.south west)+(0.25,-1.7)$){\footnotesize  \texttt{x\_min,x\_max}};
\node[anchor=north west] at($(U.south west)+(0.25,-2.0)$){\footnotesize  \texttt{q\_min,q\_max}};
\node[anchor=north west] at($(U.south west)+(0.25,-2.3)$){\footnotesize  \texttt{p\_nom,p\_std}};
%-------- methods ----
\node[anchor=north west] at($(A.south west)+(0.1,-1.7)$) (M) {\footnotesize  \color{Blue} \textsc{Methods }};
\node[anchor=north west] at($(M.south west)+(0,0)$){\footnotesize  \texttt{ode(x,u,p,q)}};
\node[anchor=north west] at($(M.south west)+(0,-0.3)$){\footnotesize  \texttt{constaints(x,u,p,q)}};
\node[anchor=north west] at($(M.south west)+(0,-0.6)$){\footnotesize  \texttt{stage\_cost(x,u,p,q)}};
\node[anchor=north west] at($(M.south west)+(0,-0.9)$){\footnotesize  \texttt{final\_penalty(x,u,p,q)}};
\node[anchor=north west] at($(M.south west)+(0,-1.2)$){\footnotesize  \texttt{generate\_cloud(n)}};
\end{tikzpicture}
\end{center} 	
\caption{\color{Blue} Summary of the main classes and functions exported by the package and the user-defined object to prepare for the a specific control problem.}\label{fig_package}
\end{figure}

\noindent In this section, a very brief description of the package is provided while a complete description can be found at the author's 
\texttt{GitHub} account (\texttt{https://github.com/mazenalamir/MPC\_tuner}). \\ \ \\ First of all, in order to use the package, the user needs to provide an instance \texttt{pb} of the \texttt{Container} class with the attributes and the methods shown in Figure \ref{fig_package}. This simply includes the different dimensions of the vectors $x, p$ and $q$, the basic sampling period $\tau$, the bounds delimiting admissible values of different vectors involved. The methods include the ODEs of the system ($f$), the definition of the constraints ($c$), the stage cost ($\ell$) the terminal penalty ($\Psi$) maps.  This object called hereafter \texttt{pb} is called by some of the functions and the classes of the package that are described hereafter. \\ \ \\ The package contains the following main classes and their main methods (Figure \ref{fig_package}): \\ \ \\
$\checkmark$ {\color{RubineRed}\underline{Class:\texttt{ Sigma}}}: an instance of this class is a specific value of $\sigma$. This class exports the maps \begin{center}
$\kappa(\alpha)$, $\texttt{N\_pred}(\alpha)$, \dots, etc. 
\end{center} which are the maps $\pi_i(\alpha)$ described in \eqref{defdepialphasigma}. The call \texttt{Sigma()} creates an instance $\sigma$ with randomly sampled properties that lie within bounds with default values although they can be chosen by the user via instantiation call.\vskip 2mm
$\checkmark$ {\color{RubineRed}\underline{Class:\texttt{ MPC}}}: an instance of this class is a specific MPC setting defined by the triplet  (\texttt{pb}, $\sigma$, $\alpha$). The main methods of this class are: \begin{center}
\texttt{MPC.feedback(x,p,q,z0}) \\
\texttt{MPC.sim\_cl(sc, z0, optim\_par)}
\end{center}  
where the \texttt{feedback} function delivers the feedback input vector while \texttt{sim\_cl} simulates the closed-loop associated to the scenario \texttt{sc} by calling \texttt{MPC.feedback} at the successive updating period using the initial guess \texttt{z0} the first time and using warm starts at the following updating steps. The computation of the success condition uses the object  \texttt{optim\_par} which incorporates the parameters $\gamma, \varepsilon$, \texttt{dev\_acc} and \texttt{c\_max} shown in Figure \ref{fig_at_a_glance} and involved in the definition of the success constraints \eqref{req1}, \eqref{req2} and \eqref{req3}. Notice that in the current version of the package, the optimization uses the framework \texttt{CasADi} \cite{Andersson2018} to define an single-shooting optimization problem to be solved using the solver \texttt{IPOPT} \cite{BIEGLER2009575}. Next version might incorporate fast gradient-based solvers that might be very computationally efficient in the context of limited number of iterations. multiple-shooting version of the current implementation can also be proposed for larger problems. \\ \ \\ 
Beside the above classes, the package contains the following main global methods:\\ \ \\
$\checkmark$ {\color{RubineRed}\underline{Function:\texttt{ Generate\_A(pb,nb,nsb)}}}: which generates a list of \texttt{nb} subsets $\mathcal A^{[\ell]}$, $\ell=1,\dots,\texttt{nb}$ of cardinality \texttt{nsb} each, leading to a total set of scenarios $\mathcal A:=\bigcup_{\ell=1}^\texttt{nb}\mathcal A^{[\ell]}$ of cardinality \begin{center}
\text{card}$(\mathcal A)$ = \texttt{nb $\times$ nsb}
\end{center} 
The first of these sets plays the role of the initial set $\mathcal A_0:=\mathcal A^{[1]}$ invoked above and used to determine the values of $\hat\alpha(\sigma^{(j)}\vert \mathcal A_0)$ to be certified using the remaining subsets $\mathcal A^{[\ell]}$, for $\ell=2,\dots, \texttt{nb}$. Note that this function admits the user-defined object \texttt{pb} as argument since the relevant method to generate relevant set of scenarios is problem-dependent. That is why the functions \texttt{Generate\_A(pb,nb,nsb)} calls the method \texttt{pb.generate\_cloud} of the user-defined object \texttt{pb} mentioned in the beginning of the present section. \\  \ \\
$\checkmark$ {\color{RubineRed}\underline{Function:\texttt{ Design\_MPC(\dots)}}}: which is the main function that enables to look for a sub-optimal NMPC setting following the algorithm described in Section \ref{sub_optimal_tractable}. The call of this main function admits the following input arguments:\vskip 1mm 
- The user defined object \texttt{pb} \vskip 1mm
- A list \texttt{S} of possible values of the shaping parameter $\sigma$ generated by successive calls (\texttt{N\_trials}) of instance generation of the class \texttt{Sigma} described above. \vskip 1mm
- The list of sets $\mathcal A:=\bigcup_{\ell=1}^{n_s}\mathcal A^{[\ell]}$ \vskip 1mm
- The object \texttt{optim\_par} described above.\vskip 1mm
This function return \texttt{Container} instance with two attributes, namely a data frame representing the set of admissible settings $(\sigma, \alpha)$ with their associated cumulated closed-loop costs and the sequence showing the total number of excluded $\sigma$-configuration during the examination of the different subsets of scenarios $\mathcal A^{[\ell]}$.

\section{An Illustrative example}\label{sec_example}
\begin{figure}[H]
\begin{center}
\begin{tikzpicture}
\draw[rounded corners, fill=Gray!10,draw=white] (-0.6,2.1) rectangle (4.2,-0.6);
\draw[<->, >=latex, thin] (0,2) -- (0,0) -- (4,0);
\coordinate (x) at(1.6,0);
\coordinate (y) at(0,0.8);
\coordinate (O) at(1.6, 0.8);
\coordinate (pl) at(0.6,0.3);
\coordinate (pr) at($(pl) ! 2 ! (O)$);
\coordinate (cl) at($(pl) ! 0.8 ! (O)$);
\coordinate (cr) at($(pl) ! 1.2 ! (O)$);
\coordinate (Al) at ($(cl)!0.17!280: (pr)$);
\coordinate (Ar) at ($(cr)!0.2!270: (pr)$);
\coordinate (Ab) at ($(O)!0.2!96: (pr)$);
\coordinate (l) at ($(O)!1.4!(pl)$);
\coordinate (r) at ($(O)!1.4!(pr)$);
\coordinate (h) at ($(O)!0.5!(pr)$);
\coordinate (K) at ($(O)+(2,0)$);
%-------
\draw[fill=Blue!10] (pl) -- (cl) -- (Al) -- (Ar) --(cr) -- (pr) -- (Ab) --cycle;
\draw[fill=Black] (1.6, 0.8) coordinate (O) circle (1pt);
\draw[dotted, thin] (x) |- (y);
\draw[dotted] (l) -- (r);
\draw[dotted] (O) -- ++(2,0);
%\draw[->] ($(O)+(0.7,0)$) arc (0:80:2.5mm) node[right,pos=0.7]{\scriptsize$\theta$};
\pic["$\theta$", draw=Brown, ->, angle eccentricity=1.1, angle radius=1.4cm, scale=1,font=\scriptsize,Brown]
    {angle=K--O--pr};
\draw[->, >=latex, MidnightBlue] (O) -- ($(O)!3!(Ab)$) node[above]{\footnotesize $u_1$};
\draw[->, >=latex, RubineRed] ($(h)+(0,-0.2)$) arc (0:90:4.5mm) node[above]{\footnotesize $u_2$};
%----
\node[below] at(x){\footnotesize $y$};
\node[left] at(y){\footnotesize $z$};
\end{tikzpicture}
\end{center} 	
\end{figure}
\begin{figure}
\begin{center}
\includegraphics[width=0.45\textwidth]{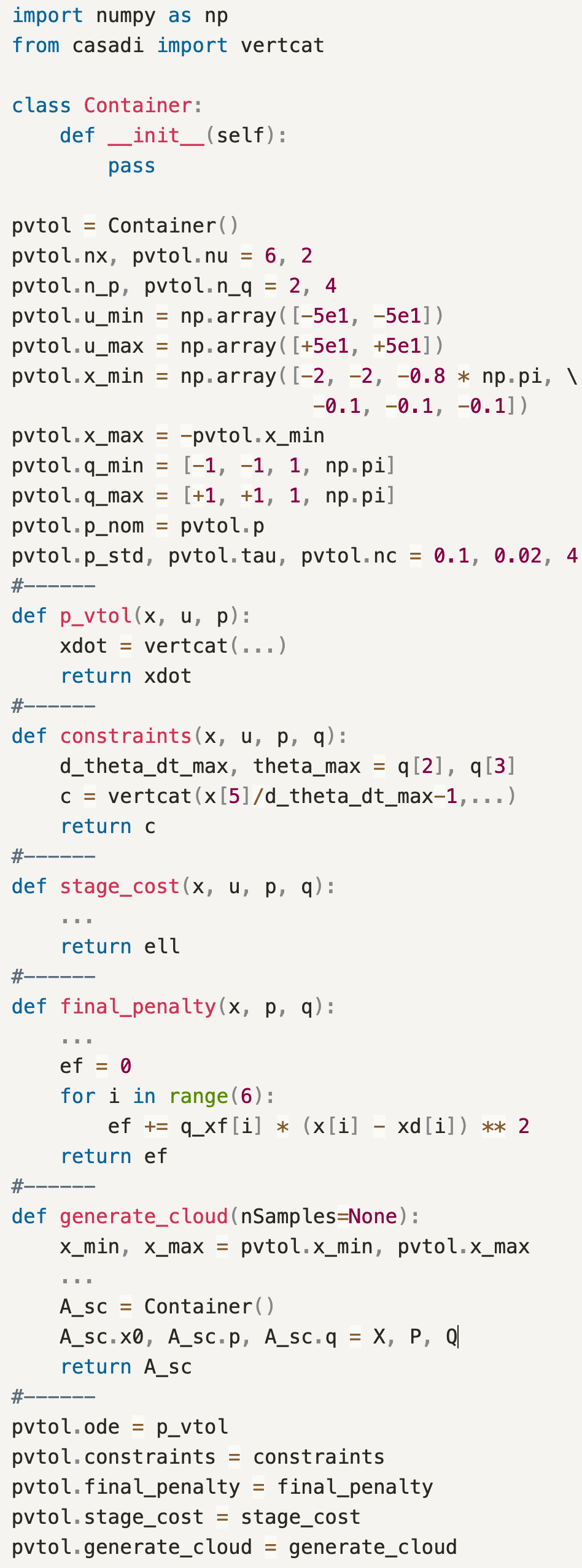}
\end{center} 
\caption{\color{Blue} The user-defined python file \texttt{user\_defined\_pvtol.py} that creates the instance \texttt{pvtol} of the \texttt{Container} class. This instance is imported by the main script file shown in Figure \ref{fig_main_script}.}\label{fig_user_defined_python}
\end{figure}
\begin{figure}
\begin{center}
\includegraphics[width=0.45\textwidth]{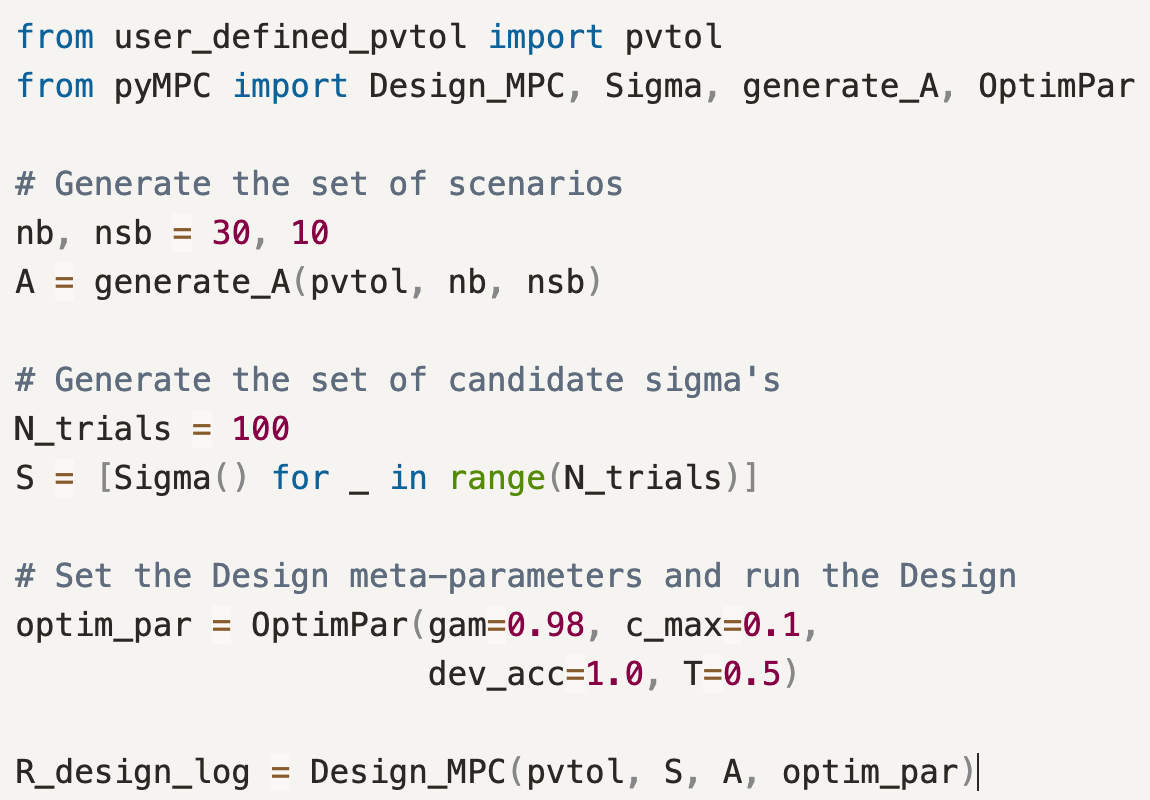}
\end{center} 
\caption{\color{Blue} The script that runs the MPC design procedure using \texttt{N\_trials}=100 candidates $\sigma$ and a set of scenarios of cardinality card($\mathcal A$) = 300 decomposed into batches $\mathcal A^{[\ell]}, \ell=1,\dots,30$ of cardinality $10$ each. Note that each experiments uses different initial set of 100 candidate configurations.}\label{fig_main_script}
\end{figure}
\begin{figure}
\begin{center}
\hrule\vskip 1mm
\texttt{dev\_acc = 1}\vskip 1mm
\hrule\vskip 1mm
\includegraphics[width=0.48\textwidth]{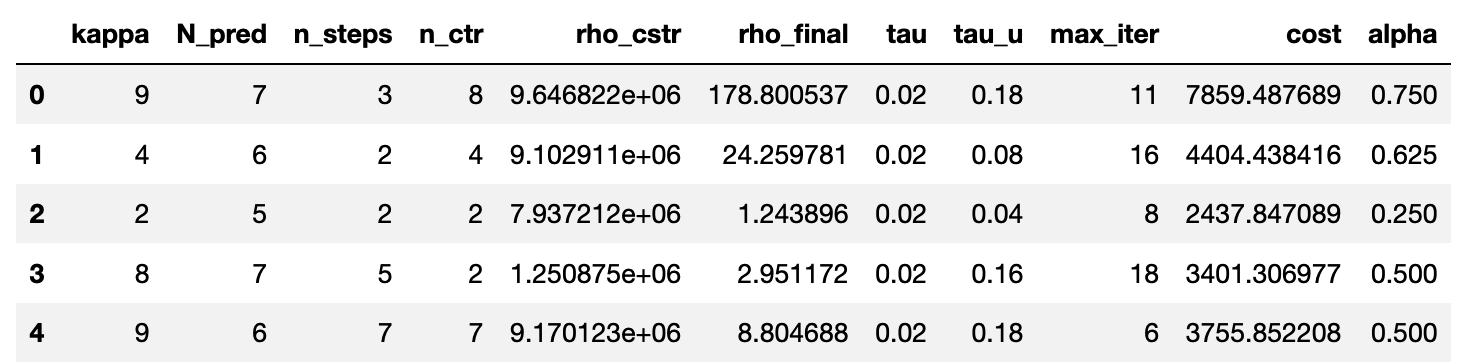} \vskip 2mm
\hrule\vskip 1mm
\texttt{dev\_acc = 2}\vskip 1mm
\hrule\vskip 1mm
\includegraphics[width=0.48\textwidth]{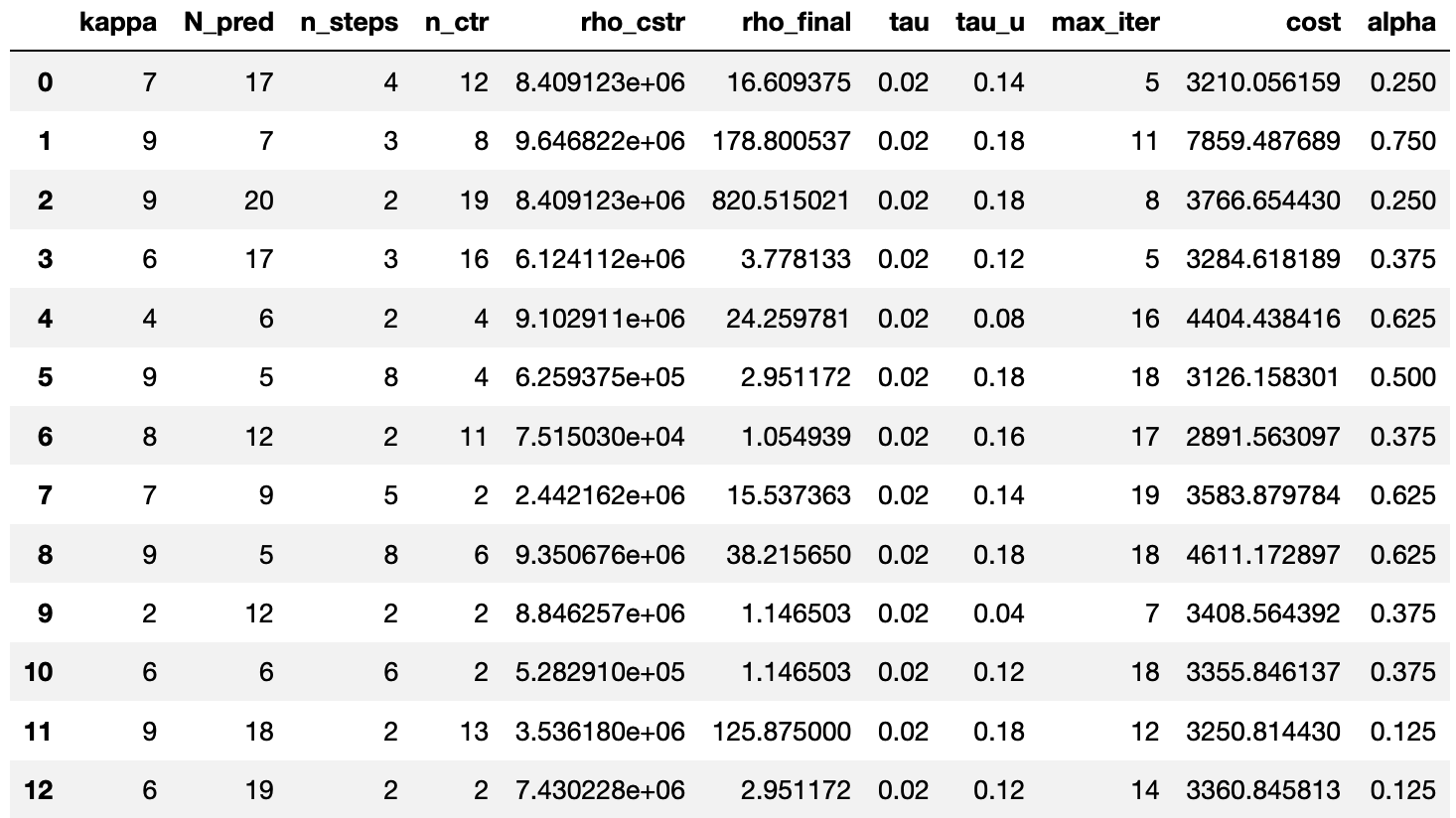} \vskip 2mm
\includegraphics[width=0.48\textwidth]{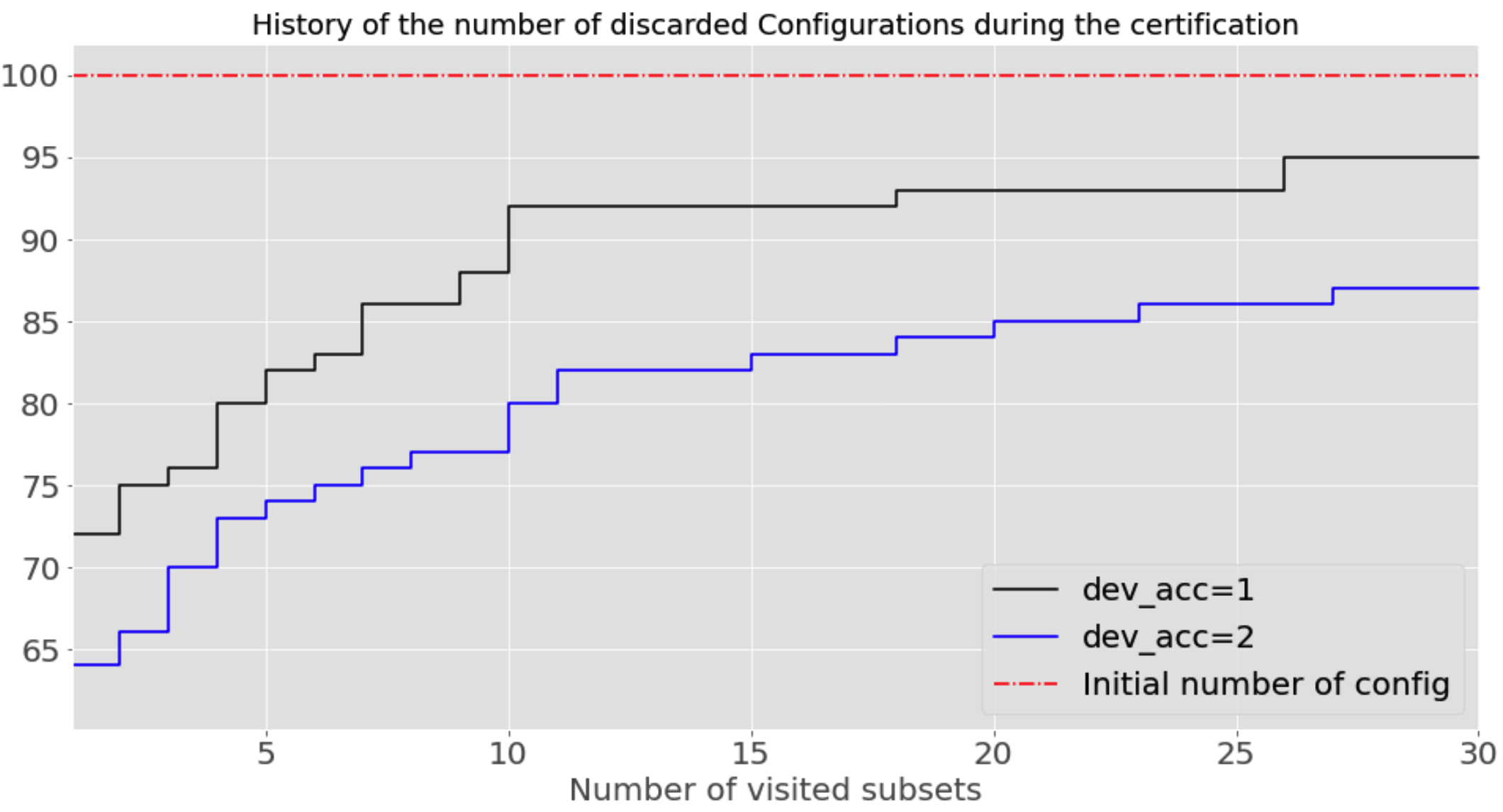}
\end{center} 
\caption{\color{Blue} Two instances of results given by the function \texttt{Design\_MPC} for different values of the \texttt{dev\_acc} input for the algorithm, namely, \texttt{dev\_acc} = 1 (upper items) and \texttt{dev\_acc=2} (lower items). In both cases, the configurations are sampled inside the domains defined by the bounds shown on Table \ref{tabBounds}.}\label{fig_results}
\end{figure}

\noindent In order to illustrate the framework and the use of the associated \texttt{python}-based package, let us consider the control of the PVTOL that obeys the following normalized dynamics:
\begin{subequations}\label{pvtol}
\begin{align}
\ddot y&=-u_1\sin\theta+p_1u_2\cos\theta \\
\ddot z&=u_1\cos\theta+p_1u_2\sin\theta-1\\
\ddot \theta&=p_2u_2 
\end{align} 	
\end{subequations}
which shows a state vector $x:=(y,z,\theta,\dot y,\dot z,\dot\theta)$ of dimension $n_x=6$, a control vector of dimension $n_u=2$. The model depends on two dimensional parameter vector $p$ ($n_p=2$). The control objective is to regulate around reference values of $y$ and $z$ that are denoted hereafter by $q_1$ and $q_2$ respectively (these are the first two components of the context vector $q$ mentioned in the precious sections. This leads to the desired targeted state $x_d:=(q_1,q_2,0,\dots,0)$ corresponding to the steady control $u_d:=(1,0)$. Therefore the economic stage cost is given by $\ell(x,u,p,q):=\|x-x_d\|_Q^2+\|u-u_d\|_R^2$ with predefined $Q$ and $R$ weighting matrices\footnote{In the numerical investigation, $Q=\text{diag}(10^3,10^3,10^3,1,1,1)$ and $R=\text{diag}(0.1,0.1)$} which are not to be tuned. On the contrary, the terminal penalty function defined by $\Psi(x,p,q):= \rho_f \|x-x_d\|_Q$ does involve the tunable parameter $\pi_5:=\rho_f$ [see \eqref{defdepi}]. All the scenarios considered hereafter last a duration of 0.5 sec which induces different number of optimization depending on the values of the parameters that induce different values of $\tau_u$. The contraction rate $\gamma=0.98$ and the threshold $c_{max}=0.1$ are used. The precision on $\alpha$ in the dichotomic search is fixed to \texttt{optim\_par.eps}$=0.15$. \\ \ \\
As for the constraints, beside the input saturation defined by $u\in [-50,+50]^2$, the following two constraints are imposed: $\vert\dot\theta\vert\le q_3$ and $\vert\theta\vert\le q_4$ which defines two other components of the context vector $q$ (hence $n_q=4$). This defines the constraints map $c(x,u,p,q)$ that encodes $n_c=4$ constraints. \\ \ \\
Figure \ref{fig_user_defined_python} shows the python file that defines the user-defined items needed for the NMPC design. Notice that not all the details are shown since many obvious script comes directly from the definition and the equations above. However, the following comments are worth giving in order to highlight some important features, namely:\vskip 1mm 
{\color{Blue} 
$\checkmark$ The \texttt{Container} class is simply a void class that accepts new attributes to be added on the flow. Notice that the object \texttt{pvtol} plays the role of the object \texttt{pb} used in the previous discussion.\vskip 1mm
$\checkmark$ Notice the systematic use of the \texttt{CasADi vertcat} command that is used to define vectors. This is mandatory since the  \texttt{CasADi} framework is used here to solve the optimization problem.  \vskip 1mm
$\checkmark$ The definition of the bounds on $q$ enables different set-points to be explored inside $[-1,+1]^2$. However, $q_3$ and $q_4$ are taken constant here meaning that the bounds on $\theta$ and $\dot\theta$ show no reason to be modified for this problem. \vskip 1mm
$\checkmark$ Notice that this script defines the nominal value \texttt{p\_nom} of the parameter vector together with its vector of standard deviations \texttt{p\_std}. This information, together with the bounds on $x$ and $q$ defined in the beginning of the script is used in the last function \texttt{generate\_cloud} in order to generate representative set of scenarios. For instance \texttt{A\_sc.x0[i]} and  \texttt{A\_sc.p[i]} provides the initial state $x_0$ and the vector of parameters $p$ for the scenario number \texttt{i}.
}
\ \\ \ \\
The object \texttt{pvtol} created by the script shown in Figure \ref{fig_user_defined_python} is imported in the main script shown in Figure \ref{fig_main_script}. This script starts by using the function \texttt{generate\_A} to create a set $\mathcal A$ of scenarios with cardinality 300 that is decomposed into \texttt{nb} = 30 batches of \texttt{nsb} = 10 scenarios each. Then the script creates a set of \texttt{N\_trials} = 100 candidate shaping parameter vectors $\sigma^{[j]}, j=1,\dots,\texttt{N\_trials}$. 
\ \\ Notice that the chosen values of \texttt{nb} and \texttt{nsb} induces an initial set of scenarios $\mathcal A_0$ with cardinality $n_0=10$ and a number of certification scenarios equal to $290$ which, according to the first line of Table \ref{lesNtab}, is \hi{sufficient to certify the successful design for a precision level of $\eta=0.05$ and a confidence level of $\delta=10^{-3}$}. \\ \ \\
\begin{table}[H]
\begin{center}
\begin{tabular}{lcc} \hline
    Design parameter & min-value & max-value\\ \hline
    \texttt{N\_pred}  & 5 & 25  \\
    $\kappa$  & 1 & 10  \\
    $\rho_f$  & 1 & $10^3$ \\
    \texttt{rho\_cstr}  & $10^3$ & $10^7$ \\
    \texttt{max\_iter}  & 5 & 20\\
 \hline
\end{tabular}
\end{center} 
\ \\
\caption{\color{Blue} The default bounds used on the different design parameters when randomly sampling the shape configuration parameter $\sigma$ via the instantiation call \texttt{Sigma()}.} \label{tabBounds} 
\end{table}
\centerline{\sc Discussion}\ \\
\noindent  Now in order to better examine the following results and assess the relevance of the proposed framework, Table \ref{tabBounds} shows the defaults bounds used in the random sampling of the shaping parameter $\sigma$ when using the instantiation call \texttt{Sigma()}. For instance when generating \texttt{N\_trials} = 100 configurations, one gets a list of $100$ uniformly randomly sampled terminal penalty parameter $\rho_f$ in the interval [1, 1000]. Notice that the call \texttt{Sigma(rho\_final\_min=10, rho\_final\_max=100,\dots)} instead of simply \texttt{Sigma()} in the script of Figure \ref{fig_main_script} enables the user to choose the bounds of all the design parameters shown in Table \ref{tabBounds} and some few other that are skipped in this presentation. \\ \ \\ Figure \ref{fig_results} shows the tuning results for two different target devices defined by \texttt{dev\_acc=1} (top) and \texttt{dev\_acc=2} (bottom) for the same randomly generated set of scenarios $\mathcal A$ and candidate set of parameters $\sigma^{(j)}, j=1,\dots, \texttt{N\_trials}$. For each case, two results are shown, namely, the data frame showing the admissible setting together with their individual parameters as well as the cumulative closed-loop cost over the scenarios contained in the set $\mathcal A$ of $300$ scenarios. Below this data frame, the evolution of the number of excluded candidate values of $\sigma$ among the \texttt{N\_trials}=100 randomly sampled design configurations is shown as a function of the number of already executed batches corresponding to the subsets of scenarios $\mathcal A^{[\ell]}$. Regarding the results, the following observations are worth making:\vskip 2mm 

\noindent$\checkmark$ \hi{Small portion of configurations is eligible}: Despite the quite reasonable bounds given in Table \ref{tabBounds}, it is quite remarkable that only a small portion of the randomly sampled settings are admissible (5\% for a computation target given by \texttt{dev\_acc=1}) and (13\% for a computation target given by \texttt{dev\_acc=2}). This simple fact suggests that the problem addressed here is quite relevant. \vskip 2mm
\noindent$\checkmark$ \hi{High values of $\rho_f$ lead to unfeasibility}: The results suggests that high values of $\rho_f$ lead to inadmissible settings and/or high values of the cost. Recall that the reported closed-loop costs do not incorporate the terminal penalty as the latter is generally used for stability reason. However, this cost includes the possible non vanishing terms coming from the constraint being violated by less than the authorized threshold $c_{max}\neq 0$ . The results suggest that too high terminal penalty might lead to bad performance when the number of iterations is limited. This is intuitively sound because in this case, the problem is stiffer and the step size is consequently small. \vskip 2mm
\noindent$\checkmark$ \hi{High values of \texttt{rho\_cstr} are needed}: On the contrary, the high majority of admissible values of soft constraints penalty \texttt{rho\_cstr} seem to lie exclusively close to the upper bound of the admissible interval $[10^3, 10^7]$. This is intuitively quite tricky to guess and enforces, if still needed, the relevance of the problem addressed in this contribution. This also suggests that it might be interesting to re-run the algorithm with the bounds of  \texttt{rho\_cstr} shifted towards higher values, for instance $[10^5, 10^9]$ following the conjecture according to which, given the random sampling, too many sampled settings fail in meeting the constraints with low values of this parameter which penalizes the randomly generated set of configurations.\vskip 2mm
\noindent$\checkmark$ \hi{The scenarios are constraints-challenging}: The previous fact also suggests that the set of scenarios used in the certification does involve constraint-violation-risky initial conditions that have been managed using high penalty on the soft constraints, otherwise, we would have successful settings with lower values of $\rho_\text{constr}$ since the latter would have no effect on the success/failure status.\vskip 2mm
\noindent$\checkmark$ \hi{Impact of \texttt{dev\_acc}}t: Although one might reasonably expect that when \texttt{dev\_acc}$>1$, the number of admissible configurations increases (which is the case in our experiments), it is not necessarily true that the best sub-optimal solution is always better since the initial randomly sampled set of candidate configurations is not the same. \vskip 1mm
\noindent$\checkmark$ \hi{Large eligible possibilities for the prediction horizon}: Notice that among the set of admissible settings, the prediction horizon lengths take values from 0.2 up to 1.26 when \texttt{dev\_acc}$=1$ and from 0.48 to 3.24 when \texttt{dev\_acc}=2. This can be explained by the definition of $\hat\alpha(\sigma\vert \mathcal A)$ being the maximum allowable value since this definition enhance longer prediction horizons \texttt{N\_pred} $\times$ \texttt{tau\_u}. The final choice of the NMPC design might favor not too short prediction horizon for obvious reasons even if this corresponds to slightly higher closed-loop since the connection between the truly obtained closed-loop performances and the ones predicted on the short term simulation used in the algorithm is not so strong as it is widely accepted by NMPC practitioners. \vskip 2mm
\noindent Obviously, the higher the cardinality of the set of candidates $\sigma$ is, the higher is the probability to get closer to a truly optimal design. This is a matter of computation time. The computation times for the design examples lie around 1 hour each on a \texttt{MacBook Pro, 2.4 GHz Intel Core i9}. Keep in mind however that the computation time depends on the randomly sampled design settings as this impact the number of problem solutions through the \texttt{N\_pred} parameter and also the number of early discarding of design settings which depends on the quality of the initial set of candidates.

\section{Conclusion \& ongoing investigation}\label{sec_conclusion}
\noindent In this paper a systematic approach and an associated freely available (\texttt{MPC\_tuner})\texttt{python}-package are proposed for the design of the \hi{implementation parameters of an NMPC controller}. Despite encouraging preliminary results, it might be conjectured that additional/different tricks might be used to accelerate the computation and or reduce the level of sub-optimality. One option would be to use Machine Learning tools to derive preliminary \underline{\textit{\color{Blue} feasibility predictors}} that can be trained over a cloud of $(\sigma,\alpha)$ without the dichotomic search used in in a first step which greatly impacts the computation time, the resulting model can then be used to make better guesses reducing hence the number of useless randomly generated samples. On the other hand, additional options can be added as part of the NMPC setting possibilities to be chosen by the algorithm. For instance, the fast-gradient approach can be added as a possible option. Different solvers inside the present \texttt{CasADi}-framework can be included in the search instead of the only \texttt{IPopT} that is currently used by default which prevent the efficiency of the warm start trick in reducing the simulation time for a single closed-loop experiment. 
%}
% References
\bibliographystyle{IEEEtran}
\bibliography{bibMPC.bib}  % Add your bibliography file here

\begin{thebibliography}{10}
\providecommand{\url}[1]{#1}
\csname url@samestyle\endcsname
\providecommand{\newblock}{\relax}
\providecommand{\bibinfo}[2]{#2}
\providecommand{\BIBentrySTDinterwordspacing}{\spaceskip=0pt\relax}
\providecommand{\BIBentryALTinterwordstretchfactor}{4}
\providecommand{\BIBentryALTinterwordspacing}{\spaceskip=\fontdimen2\font plus
\BIBentryALTinterwordstretchfactor\fontdimen3\font minus
  \fontdimen4\font\relax}
\providecommand{\BIBforeignlanguage}[2]{{%
\expandafter\ifx\csname l@#1\endcsname\relax
\typeout{** WARNING: IEEEtran.bst: No hyphenation pattern has been}%
\typeout{** loaded for the language `#1'. Using the pattern for}%
\typeout{** the default language instead.}%
\else
\language=\csname l@#1\endcsname
\fi
#2}}
\providecommand{\BIBdecl}{\relax}
\BIBdecl

\bibitem{Mayne2000}
D.~Q. Mayne, J.~Rawlings, C.~V. Rao, and P.~O.~M. Scokaert, ``Constrained model
  predictive control: Stability and optimality,'' \emph{Automatica}, vol.~36,
  pp. 789--814, 2000.

\bibitem{rawlings2017model}
J.~B. Rawlings, D.~Q. Mayne, and M.~Diehl, \emph{Model predictive control:
  theory, computation, and design}.\hskip 1em plus 0.5em minus 0.4em\relax Nob
  Hill Publishing Madison, WI, 2017, vol.~2.

\bibitem{Andersson2018}
J.~A.~E. Andersson, J.~Gillis, G.~Horn, J.~B. Rawlings, and M.~Diehl,
  ``{CasADi} -- {A} software framework for nonlinear optimization and optimal
  control,'' \emph{Mathematical Programming Computation}, 2018.

\bibitem{biegler2009large}
L.~T. Biegler and V.~M. Zavala, ``Large-scale nonlinear programming using
  ipopt: An integrating framework for enterprise-wide dynamic optimization,''
  \emph{Computers \& Chemical Engineering}, vol.~33, no.~3, pp. 575--582, 2009.

\bibitem{houska2011acado}
B.~Houska, H.~J. Ferreau, and M.~Diehl, ``Acado toolkit: An open-source
  framework for automatic control and dynamic optimization,'' \emph{Optimal
  Control Applications and Methods}, vol.~32, no.~3, pp. 298--312, 2011.

\bibitem{alamir2001nonlinear}
M.~Alamir, ``Nonlinear receding horizon sub-optimal guidance law for the
  minimum interception time problem,'' \emph{Control Engineering Practice},
  vol.~9, no.~1, pp. 107--116, 2001.

\bibitem{diehl2002real}
M.~Diehl, H.~G. Bock, J.~P. Schl{\"o}der, R.~Findeisen, Z.~Nagy, and
  F.~Allg{\"o}wer, ``Real-time optimization and nonlinear model predictive
  control of processes governed by differential-algebraic equations,''
  \emph{Journal of Process Control}, vol.~12, no.~4, pp. 577--585, 2002.

\bibitem{diehl2005real}
M.~Diehl, H.~G. Bock, and J.~P. Schl{\"o}der, ``A real-time iteration scheme
  for nonlinear optimization in optimal feedback control,'' \emph{SIAM Journal
  on control and optimization}, vol.~43, no.~5, pp. 1714--1736, 2005.

\bibitem{gros2020linear}
S.~Gros, M.~Zanon, R.~Quirynen, A.~Bemporad, and M.~Diehl, ``From linear to
  nonlinear mpc: bridging the gap via the real-time iteration,''
  \emph{International Journal of Control}, vol.~93, no.~1, pp. 62--80, 2020.

\bibitem{alamir2009framework}
M.~Alamir, ``A framework for monitoring control updating period in real-time
  {NMPC} schemes,'' \emph{Nonlinear Model Predictive Control: Towards New
  Challenging Applications}, pp. 433--445, 2009.

\bibitem{alamir2013monitoring}
------, ``Monitoring control updating period in fast gradient based {NMPC},''
  in \emph{2013 European Control Conference (ECC)}.\hskip 1em plus 0.5em minus
  0.4em\relax IEEE, 2013, pp. 3621--3626.

\bibitem{alamir2016state}
------, ``A state-dependent updating period for certified real-time model
  predictive control,'' \emph{IEEE Transactions on Automatic Control}, vol.~62,
  no.~5, pp. 2464--2469, 2016.

\bibitem{bonne2017experimental}
F.~Bonne, M.~Alamir, and P.~Bonnay, ``Experimental investigation of control
  updating period monitoring in industrial {PLC}-based fast {MPC}: Application
  to the constrained control of a cryogenic refrigerator,'' \emph{Control
  Theory and Technology}, vol.~15, no.~2, pp. 92--108, 2017.

\bibitem{alamir2015certification}
M.~Alamir, ``From certification of algorithms to certified mpc: The missing
  links,'' \emph{IFAC-PapersOnLine}, vol.~48, no.~23, pp. 65--72, 2015.

\bibitem{richter2011computational}
S.~Richter, C.~N. Jones, and M.~Morari, ``Computational complexity
  certification for real-time mpc with input constraints based on the fast
  gradient method,'' \emph{IEEE Transactions on Automatic Control}, vol.~57,
  no.~6, pp. 1391--1403, 2011.

\bibitem{pu2016complexity}
Y.~Pu, M.~N. Zeilinger, and C.~N. Jones, ``Complexity certification of the fast
  alternating minimization algorithm for linear mpc,'' \emph{IEEE Transactions
  on Automatic Control}, vol.~62, no.~2, pp. 888--893, 2016.

\bibitem{alamir2006stabilization}
M.~Alamir, \emph{Stabilization of nonlinear systems using receding-horizon
  control schemes: a parametrized approach for fast systems}.\hskip 1em plus
  0.5em minus 0.4em\relax Springer, 2006, vol. 339.

\bibitem{alamir2017contraction}
------, ``Contraction-based nonlinear model predictive control formulation
  without stability-related terminal constraints,'' \emph{Automatica}, vol.~75,
  pp. 288--292, 2017.

\bibitem{alamo2009}
T.~Alamo, R.~Tempo, and E.~Camacho, ``Randomized strategies for probabilistic
  solutions of uncertain feasibility and optimization problems,''
  \emph{Automatic Control, IEEE Transactions on}, vol.~54, no.~11, pp.
  2545--2559, Nov 2009.

\bibitem{BIEGLER2009575}
\BIBentryALTinterwordspacing
L.~Biegler and V.~Zavala, ``Large-scale nonlinear programming using ipopt: An
  integrating framework for enterprise-wide dynamic optimization,''
  \emph{Computers \& Chemical Engineering}, vol.~33, no.~3, pp. 575--582, 2009,
  selected Papers from the 17th European Symposium on Computer Aided Process
  Engineering held in Bucharest, Romania, May 2007. [Online]. Available:
  \url{https://www.sciencedirect.com/science/article/pii/S0098135408001646}
\BIBentrySTDinterwordspacing

\end{thebibliography}

\end{document}